\documentclass[Afour,sageh,times]{sagej}

\usepackage{moreverb,url}

\usepackage[colorlinks,bookmarksopen,bookmarksnumbered,citecolor=red,urlcolor=red]{hyperref}

\newcommand\BibTeX{{\rmfamily B\kern-.05em \textsc{i\kern-.025em b}\kern-.08em
T\kern-.1667em\lower.7ex\hbox{E}\kern-.125emX}}

\def\volumeyear{2025}

\begin{document}

\runninghead{Bäumer et. al}

\title{Evaluation of an ITD-to-ILD transformation as a method to restore the spatial benefit in speech intelligibility in hearing impaired listeners}

\author{Timm-Jonas Bäumer\affilnum{1}, Johannes W. de Vries\affilnum{2}, Stephan Töpken\affilnum{1}, Richard C. Hendriks\affilnum{2}, Peyman Goli\affilnum{1} and Steven van de Par\affilnum{1}}

\affiliation{\affilnum{1}Dept. of Med. Physics and Acoustics, Cluster of Excellence Hearing4all, Carl von Ossietzky Universität Oldenburg, Oldenburg, 26129, Germany\\
\affilnum{2}Faculty of Electrical Engineering, Mathematics and Computer Science, Delft University of Technology, Delft, 2628, the Netherlands}

\corrauth{Timm-Jonas Bäumer, Carl von Ossietzky Universität Oldenburg
Acoustics Group, Carl-von-Ossietzky-Straße 9-11, 26129 Oldenburg, Germany}

\email{timm-jonas.baeumer@uni-oldenburg.de}

\begin{abstract}

To improve speech intelligibility in complex everyday situations, the human auditory system partially relies on Interaural Time Differences (ITDs) and Interaural Level Differences (ILDs). However, hearing impaired (HI) listeners often exhibit limited sensitivity to ITDs, resulting in decreased speech intelligibility performance. This study aimed to investigate whether transforming low-frequency ITDs into ILDs could reintroduce a binaural benefit for HI listeners.

We conducted two experiments with HI listeners. The first experiment used binaurally phase-shifted sinusoids at different frequencies to evaluate the HI listeners ITD sensitivity threshold. All subjects had an increased ITD threshold at higher frequencies, with different ITD sensitivities between the subjects in the lower frequencies. In the second experiment, Speech Reception Thresholds (SRTs) were measured in different binaural configurations by manipulating Head-Related Transfer Functions (HRTFs). The results showed that, despite the decreased ITD sensitivity, removing ITDs decreased SRTs by approximately 1 dB compared to the unprocessed baseline, where ITDs and ILDs are available. Furthermore, substituting low-frequency ITDs with ILDs yielded an improvement for a lateral target speaker. Adding the low-frequency ILDs while preserving the ITDs caused a significant improvement for speakers in all directions. 

These findings suggest that the proposed transformation method could be effective in restoring binaural benefits in HI listeners. The results of this study suggest the use of such transformation techniques to be implemented in hearing aids and cochlear implants, directly benefiting HI listeners.

\end{abstract}


\maketitle

\section{Introduction}

One of the key functions of the human auditory system is the ability to interpret interaural cues, based on the signals the left and right ears receive. The two main interaural cues are the Interaural Level Differences (ILDs) and Interaural Time Differences (ITDs). Normal Hearing (NH) listeners make use of both ITDs and ILDs to localize sound sources (\cite{Blauert1997}). It is generally accepted that ITDs are dominant at low frequencies ($< 1500$ Hz), while ILDs are dominant at higher frequencies ($> 1500$ Hz)  (e.g. \cite{Hartmann2016}). In acoustically challenging situations, such as the well-known cocktail-party scenario (\cite{Cherry_1953}), the interaural cues are able to improve our ability to understand a desired speaker even though interfering speakers are present, as long as target and interferers are spatially separated. This effect is called spatial release from masking (SRM) (e.g. \cite{Doll1995}). The three main underlying mechanisms that contribute to SRM are binaural unmasking, better-ear listening and auditory stream segregation (\cite{Schoenmaker2016, Culling2021}). Binaural unmasking refers to an improvement in the internal signal to noise ratio (SNR) by aligning and subtracting the interfering signal from the target signal based on the signals of both ears, in accordance with the Equalization-Cancellation (EC) theory \citep{Durlach1963}. Some auditory models, such as the BSIM \citep{Beutelmann2006, Beutelmann2010, Rennies2011} or the binaural STOI model \citep{Andersen2015}, use an EC model together with a speech intelligibility measure to predict speech intelligibility in different scenarios. 
Better ear listening refers to the auditory system's ability to choose the ear which has the better SNR at a certain moment in time, due to, for example, gaps in one of the interferer's speech placed at one side of the listener\citep{Bronkhorst1988}. 
Auditory stream segregation is used to select only the elements of the target speaker on which attention is focussed using their common spatial and spectral characteristics (e.g. \cite{Cho2022}) to help improve speech intelligibility. All of these mechanisms benefit from the availability of the interaural cues.

Hearing impaired (HI) and elderly listeners, however, often have a reduced sensitivity to ITDs (e.g. \cite{Best2022}). Many different factors can contribute to the ITD perception deficit. In the low-frequency region, the loss in temporal fine structure (TFS) sensitivity is assumed to be the main cause for reduced ITD perception. This could be due to the reduced number of auditory nerve fibers in damaged cochleas (\cite{Schuknecht1955}), unusual phase responses on the basilar membrane as a result of hearing impairment (\cite{Ruggero1996}), poorer frequency selectivity (\cite{Hopkins2011}) or a decrease in phase locking, in which the TFS information is encoded (\cite{Füllgrabe2017}).
This decrease in ITD sensitivity results in decreased speech-intelligibility performance in challenging situations due to a less effective binaural unmasking and auditory stream segregation. HI listeners with reduced ITD sensitivity, however, can still be sensitive to ILDs, with sensitivities similar to a normal hearing (NH) control group (\cite{Hawkins1980}). ILDs can be easily transmitted via hearing aids and cochlear implants by adjusting the gains in different frequency bands. For most binaurally implanted Cochlear Implants (BiCIs), due to mismatches in the placement of electrodes, asynchronous processors and lack of TFS encoding, ITDs are not available.

One option to counteract this problem would be to transform the ITDs into low-frequency ILDs, which could serve as a cue to improve auditory stream segregation, and may also support better ear listening. A few of such transformation methods have been proposed by other researchers in the past. 
\cite{Brown2014} implemented an ITD-to-ILD transformation for BiCI users, where ITDs of up to 600 µs were converted to ILDs of $\pm$30 dB. The ILD enhancement was then applied to 20 Equivalent Rectangular Bands (ERBs) below 2~kHz, by decreasing the level on the contralateral ear. He found a large improvement in speech intelligibility in a single masker situation, when comparing the cue enhancement to the unprocessed case. In a follow-up study (\cite{Brown2018}), he also showed a significant improvement in the subjects localization performance when applying this method to only six one octave wide bands. In \cite{Richardson2025}, the transformation was extended to a higher frequency range (up to 5.5~kHz), divided into only four frequency bands in total. Albeit less beneficial than in \cite{Brown2014}, experiments with NH listeners with vocoded stimuli and HI listeners with BiCIs still showed a substantial improvement in speech intelligibility.

In a different approach, \cite{Gajecki2021} used an ILD enhancement for BiCIs by attenuating the contralateral side between 1 kHz and 10 kHz based on the sounds direction of arrival (DOA). This approach focuses on the high-frequencies instead of on the low frequencies and resulted in an improvement in the subject's localization performance.

\cite{Calis2021} presented a low-frequency ILD enhancement method using a beamformer, where existing low frequency ILDs below 800 Hz were magnified using a scaling factor. While this approach yielded a benefit regarding localization in anechoic environments, such an approach could possible be less robust due to errors when the underlying ILD that is magnified varies.

Recently, \cite{Bäumer2025} proposed a simple method to transform ITDs into low-frequency ILDs, where it was investigated how much speech intelligibility improvement can be gained when NH listeners were presented with low-frequency ILDs in the absence of ITDs, along with other binaural cue manipulations. For NH listeners in an anechoic environment, an overall SRM of 7.1 dB was found when target and interferers were spatially separated, a reduction in SRTs of 3.4 dB when ITDs were not available (to simulate severe spatial hearing loss), and a reduction of only 0.4~dB when ITDs were unavailable, but transformed to low-frequency ILDs, meaning the normal hearing capabilities were essentially restored. For a lateral target speaker with a central and contralateral interferer, SRTs even improved by 1.9 dB over the NH baseline due to the transformation. The present study investigates, whether subjects with an actual hearing impairment would also benefit from this cue transformation method, and whether the simple removal of all ITDs was an accurate approach to simulate spatial hearing loss for NH listeners. To address these questions, we repeated the same experiment as in \cite{Bäumer2025} with HI listeners and also measured their sensitivity to ITDs in this work. Furthermore, an additional condition was tested, where the low-frequency ILDs are added without removing the ITDs beforehand. This resembles more closely what a hearing aid algorithm would do, and would possibly improve performance even further since the listeners can make use of any residual ITD processing.
  
\section{\label{sec:2} Methods}
In this work, speech reception thresholds (SRTs) are measured for different spatial configurations of target and interfering speakers. To investigate the effect of several different binaural conditions such as transforming ITDs to ILDs, all necessary manipulations were done on the HRTF level. These manipulated HRTFs were later used in the listening experiment. For this, HRTFs of an artificial head (KEMAR) measured in Aachen \citep{Braren2020} were used, after resampling them to 44.1~kHz.

\subsection{\label{ch:transformationMethod} ITD-to-ILD transformation method}
This section describes the binaural cue conversion method used in this study (based on \cite{Bäumer2025}), transforming ITDs to low-frequency ILDs.

\begin{figure}[ht]
\begin{center}
\includegraphics[width = .45\textwidth]{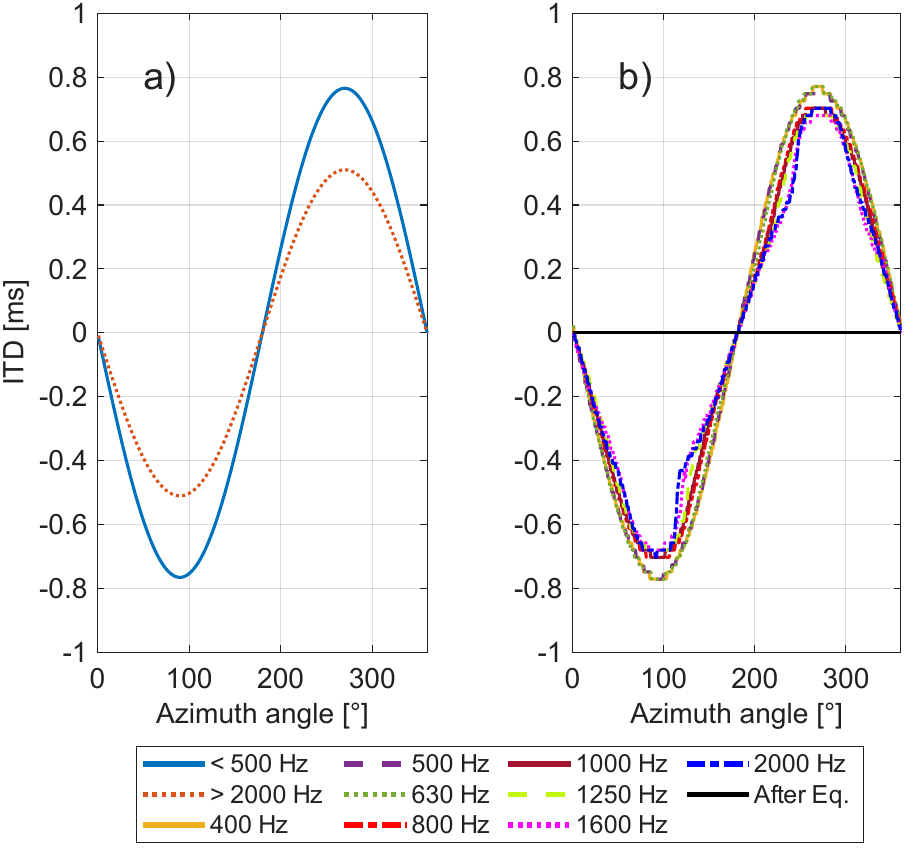}
\caption{\label{fig:ITDs}{Estimation of ITD values in third-octave bands. Left side (a) shows the theoretically expected values for high and low frequencies based on the model of \citet{Kuhn1977}. Right side (b) shows the ITDs estimated from the HRTFs per third-octave band, as well as the removed ITDs after the equalization of the IPD between left and right (black line).}}
\end{center}
\end{figure}

In the first step, the ITDs within the HRTF dataset were estimated by calculating the delay of the interaural cross-correlation's maximum between the left and right channels. Since ITDs are frequency-dependent (\cite{Otani2021}), they were estimated in third-octave frequency bands between 400 Hz and 2000 Hz. Due to the size of the loudspeakers used in \cite{Braren2020}, low frequencies sounds could not be measured reliably, and thus, the ITDs are assumed to be constant below 400 Hz (\cite{Xie2008}). Figure \ref{fig:ITDs} a) shows the theoretically expected ITDs for low ($<$~500 Hz) and high ($>$~2000 Hz) frequencies based on the model of \cite{Kuhn1977}. Figure \ref{fig:ITDs} b) shows the ITDs estimated from the HRTFs in third-octave frequency bands. 

The ITDs were then transformed to low-frequency ILDs independently for each azimuth angle $\theta$ (DOA on the horizontal plane) and frequency band $f$. The transformation was applied using Eq. \eqref{eq:transformation}
\begin{equation}\label{eq:transformation}
    \Delta \text{ILD}_{f,\theta} = 12~\text{(dB)} \times \frac{\text{ITD}_{f,\theta}}{700~\text{(µs)}},
\end{equation}
where the maximum ITD value of 700 µs and the maximum ILD value of 12 dB were chosen to be the approximately expected values for a source from 90° at 1.5 kHz (\cite{Blauert1997, Llado2022}). Below 400 Hz the ITDs and, thus, the $\Delta$ILDs were assumed to be constant (Eq. \eqref{eq:const_val})
\begin{equation}\label{eq:const_val}
    \Delta ILD_{f_{<400},\theta} = \Delta ILD_{f_{400},\theta}.
\end{equation}
$\Delta$ILD was then used to calculate the gain factor $\alpha$ per frequency band $f$ and horizontal azimuth DOA $\theta$ from Eq. \eqref{eq:alpha}. 
\begin{equation} \label{eq:alpha}
    \alpha_{f,\theta} = 10^{\Delta\text{ILD}_{f,\theta} / 20}
\end{equation}
$\alpha$ is then used to amplify the ipsilateral spectrum ($L/R$, respectively) (depending on the horizontal azimuth DOA) by half the gain and attenuate the contralateral spectrum ($R/L$, respectively) by the other half, for each frequency bin $k_f$ in the frequency band $f$ (Eq. \eqref{eq:trafo}, \eqref{eq:trafo2}), i.e.
\begin{flalign}\label{eq:trafo}
L'(k_f, \theta) = \sqrt{\alpha_{f,\theta}} \times L(k_f, \theta)\\
R'(k_f, \theta) = \frac{1}{\sqrt{\alpha_{f,\theta}}} \times R(k_f, \theta).\label{eq:trafo2}
\end{flalign}
The upper and lower frequency for each third-octave frequency band was based on \cite{ISO266}. The transformation was applied to the low-frequency part of the HRTF, up until 1000 Hz. In the area of 1000 Hz - 2000~Hz, a gradual transition between the manipulated and the original HRTF was applied to preserve the original high-frequency ILDs. Additionally, due to jumps between different gains in neighbouring frequency bands, the manipulated low-frequency spectrum was smoothed in third-octave frequency bands according to the method presented in \citet{Rasumow2014}, which did not introduce spectral colouration or audible phase-shifts.

Lastly, in order to compensate potential excessive better-ear listening cues from the spectral manipulation and to ensure comparability between different test conditions, the energy within the HRTFs after the transformation ($E'$, Eq.\eqref{eq:energy2}) was equalized to the initial energy before the transformation ($E$, Eq. \eqref{eq:energy1}). 
\begin{equation}\label{eq:energy2}
    E' = \sum_k L'^2(k) + R'^2(k).
\end{equation}
\begin{equation}\label{eq:energy1}
    E = \sum_k L^2(k) + R^2(k).
\end{equation}

This final matching step was done using Eq. \eqref{eq:energyMatch}

\begin{equation} \label{eq:energyMatch}
L''(k) = L'(k)\times\sqrt{\frac{E}{E'}}, ~~~~~~~~R''(k) = R'(k)\times\sqrt{\frac{E}{E'}}.
\end{equation}

\section{\label{sec:experimentParameters} Listening Test Methods}

\subsection{Participants}
Eight participants with mild to moderate hearing loss were selected by the Hörzentrum in Oldenburg from their database of hearing impaired people who are volunteering to participate in listening experiments. Subjects with hearing loss similar to the N3 hearing loss by \cite{Bisgaard2010} were chosen. Two of the subjects were male and six were female. The subjects were between 54 and 81 years old, with an average age of 66.9 years. 

\begin{figure}[h!]
    \centering
    \includegraphics[width=1\linewidth]{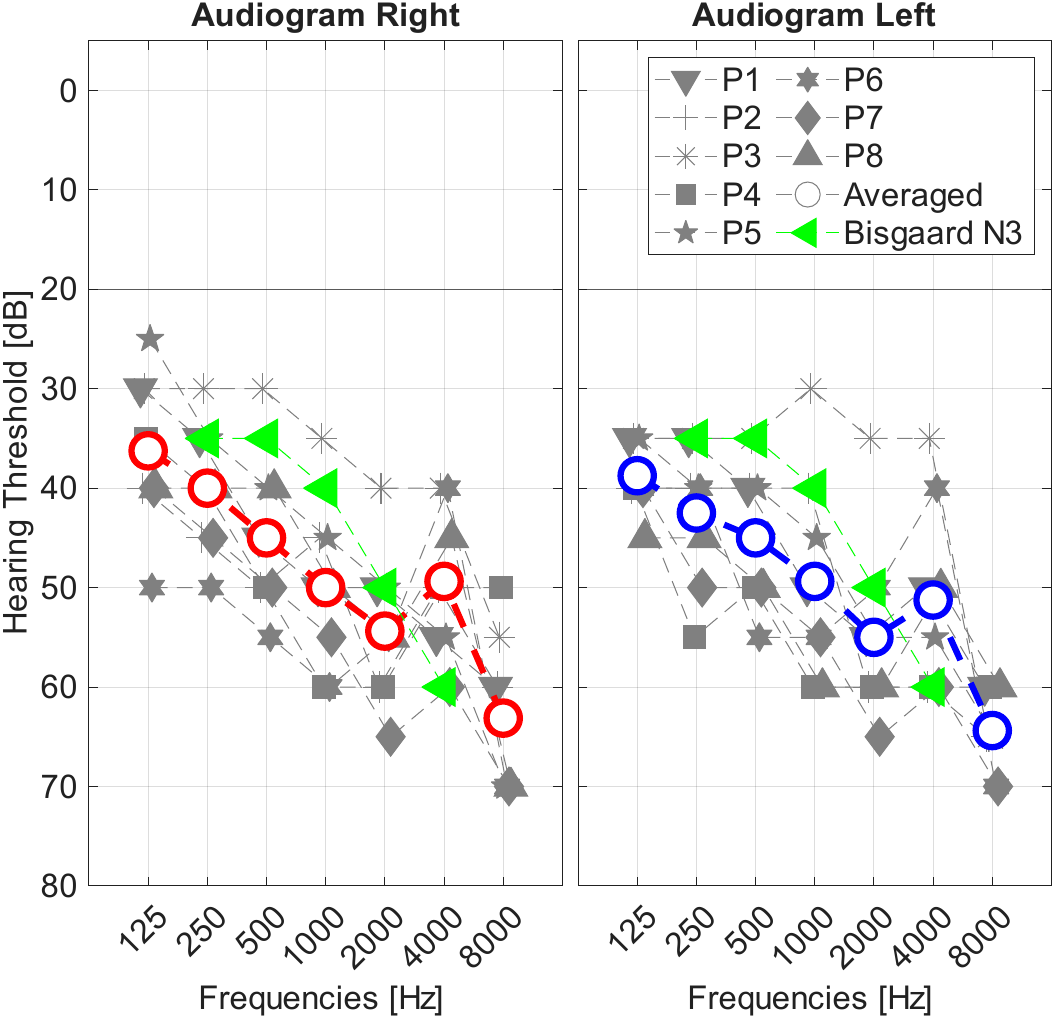}
    \caption{Individual audiograms per subject P1 to P8 (grey), average audiogram across all participants for the right (red) and left (blue) ear, and N3 hearing loss based on \cite{Bisgaard2010} (green).}
    \label{fig:audiograms}
\end{figure}

In order to avoid discrepancies with different hearing losses per side, only subjects with mostly symmetrical hearing loss (not more than 10 dB difference per frequency band between left and right) were chosen. The individual subjects' audiograms, the average audiogram across all subjects for the left and right ear, as well as the N3 line by \cite{Bisgaard2010} are shown in Fig. \ref{fig:audiograms}.

Each subject (except for P1) wears hearing aids in daily life. The experiments, however, were conducted without hearing aids. The measurements were repeated twice (three measurements in total) in 3-4 appointments of 1.5-2~h each for each participant, depending on the subject's constitution and speed. The subjects were paid 12€ per hour for their participation. 

Four of the subjects were experienced with the Oldenburg Sentence Test (OLSA, \cite{Wagener1999}) that was used in the second experiment, the others had prior experience with other speech tests such as the Freiburg monosyllabic and multisyllabic tests \citep{Hahlbrock1953}.

\subsection{\label{subsec:setup} Setup}

The experiments took place in single walled listening booths placed in a sound insulated room. The stimuli were presented to the subjects over headphones (Sennheiser HD 650) driven by the headphone output of an RME FireFace UC audio interface. Stimuli and measurement software were generated and played back via a custom measurement software in Matlab \citep{Matlab}.

\section{Experiment 1: ITD Sensitivity}\label{ch:ITDs}

To assess participants' sensitivity to ITDs, an adaptive staircase procedure was employed. In this experiment, sinusoids at five frequencies (250 Hz, 500 Hz, 750 Hz, 1000~Hz, and 1250 Hz) were presented through headphones at 75 dB SPL (calibrated with an artificial ear). ITDs were introduced by phase-shifting the sinusoidal waves.

Each trial consisted of two sinusoidal stimuli, one presented with a phase shift to the left and the other to the right, with the order randomized (i.e., either "left-right" or "right-left"). Participants were instructed to identify whether the second sound was perceived to the left or right of the first sound, following a similar procedure as described by \cite{Brughera2013}. They gave their answer by pressing a button on a GUI, and received feedback whether their answer was correct. The maximum possible ITD for each frequency, corresponding to a phase shift of $\pi$, as well as the starting ITD values, are provided in Table \ref{tab:ITDs}.

\begin{table}[htbp]
 \caption{\label{tab:ITDs} Maximum and starting ITD values for the frequencies used in the ITD pure-tone threshold experiment.}
\begin{tabular}{|c||c|c|} 
\hline
  Frequency [Hz] & Max. ITD [µs] & Starting ITD [µs]\\ [0.5ex] 
 \hline\hline
 250 & 2000 & 1000  \\ 
 \hline
 500 & 1000 & 500  \\
 \hline
 750 & 667 & 333  \\
 \hline
 1000 & 500 & 250  \\
 \hline
 1250 & 400 & 200  \\ 
 \hline
\end{tabular}
\end{table}

Each trial consisted of two signals at the specified frequency, each 500 ms in duration, where both faded-in and out with a 50 ms Hann window and a 500 ms silent interval in between them.

To familiarize the subjects with the measurement procedure and to ensure the stimuli are above hearing threshold, the first measurement started with a training phase. In the training, the sinusoids were lateralized using ILDs, by decreasing the amplitude on one side by 6~dB. At each frequency, the subjects were presented with 3-5 stimuli. After the training phase, the measurement started with the 250 Hz phase-shifted sinusoids.

To optimize the measurement process, ITDs were varied in logarithmic steps. The logarithmic transformation of ITD values was applied using the following equation (Eq. \eqref{eq:log}):

\begin{equation}\label{eq:log}
    ITD_{f,\text{log}} = 20 \times \log_{10}(ITD_{f,\text{lin}}).
\end{equation}

The logarithmic ITD values were altered in step sizes of 8, 4, 2, and 1 'dB', per the adaptive staircase method. To convert the logarithmic ITDs back to linear values, the inverse transformation (Eq. \eqref{eq:lin}) was used:

\begin{equation}\label{eq:lin}
    ITD_{f,\text{lin}} = 10^{(ITD_{f,\text{log}} / 20)}.
\end{equation}

The linear ITD values (in microseconds) were then converted into interaural phase differences (IPDs) in radians using equations Eq. \eqref{eq:lin2} and Eq. \eqref{eq:lin3}, and afterwards normalized to $2\pi$,

\begin{equation}\label{eq:lin2}
    IPD_{\phi,\text{lin}} \, [^\circ] = 360 \times f \times ITD_{f,\text{lin}} \, [\mu\text{s}]
\end{equation}
\begin{equation}\label{eq:lin3}
        IPD_{\phi,\text{lin}} \, [\text{rad}] = \frac{IPD_{\phi,\text{lin}} \, [^\circ]\times \pi}{180}
\end{equation}

The adaptive staircase procedure followed a 1-up, 2-down procedure: ITDs were reduced after two consecutive correct responses and increased following an incorrect response. This approach allowed for the determination of the ITD threshold at which participants could no longer reliably differentiate the left and right sounds.

\subsection{Results}\label{ch:ITDresults}
\begin{figure}
    \centering
    \includegraphics[width=1\linewidth]{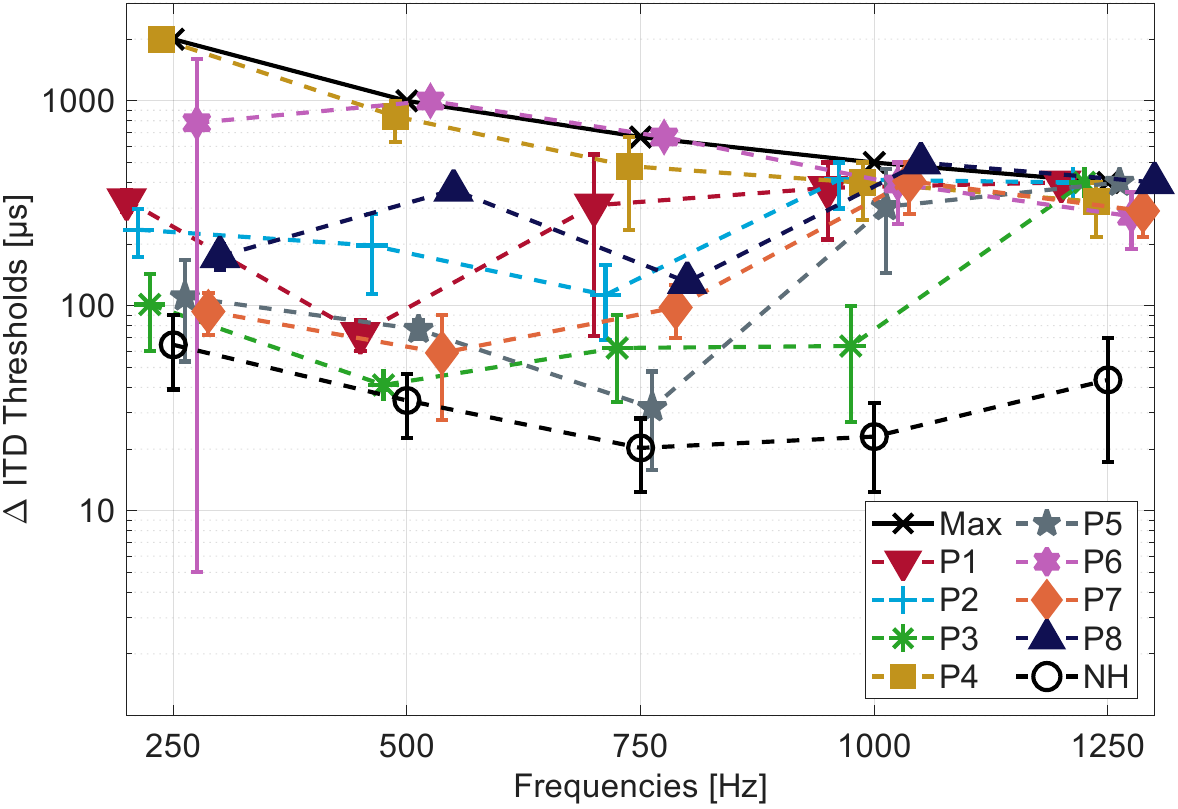}
    \caption{$\Delta$ITD thresholds of normal hearing listeners according to \cite{Brughera2013} (black circles), maximum values based on a phase shift of $\pi$ per frequency (black crosses) and thresholds of all participants P1 to P8 averaged over three measurements (coloured symbols). The errorbars indicate the mean absolute deviation.}
    \label{fig:ITD_tresholds}
\end{figure}

The averaged results of the subjects' three measurements (along with the mean absolute deviation), the maximum possible values (see Tab. \ref{tab:ITDs}) and the values for NH listeners from \cite{Brughera2013} are shown in Fig. \ref{fig:ITD_tresholds}.

The results show a large variability in ITD sensitivity between the individual subjects. For lower frequencies, P3, P5 and P7 perform the best, with only a small deviation from NH results. On the other hand, the thresholds of P4 and P6 are always close to, or at the maximum value across all frequencies, indicating a highly impaired ITD sensitivity. Towards the higher frequencies the task becomes more difficult, and all subject's thresholds approach the maximum ITD value and deviate considerably from the NH performance. This indicates that all subjects show some reduction in ITD sensitivity, and are, thus, subjects who could possibly benefit from our proposed ITD-to-ILD transformation. 

\begin{figure}[h!]
    \centering
    \includegraphics[width = 0.8\linewidth]{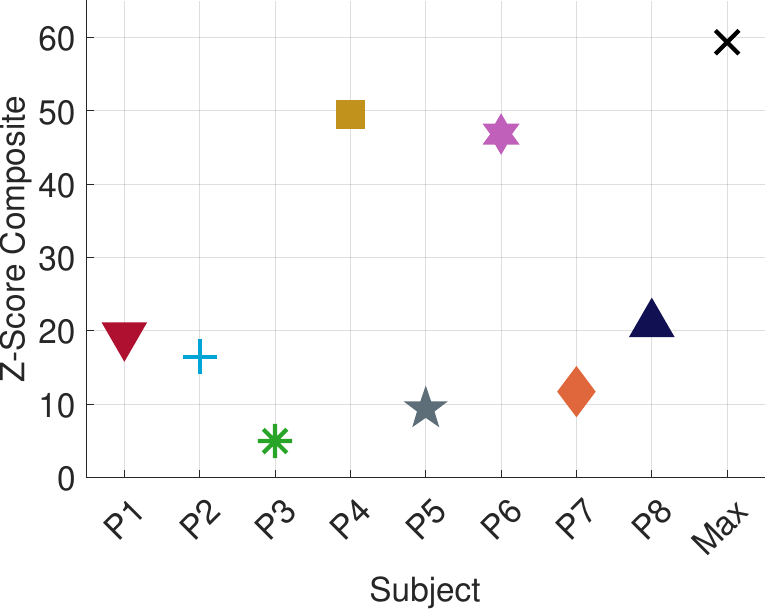}
    \caption{Composite z-scores of each hearing impaired subject based on the normal hearing performance from \cite{Brughera2013}.}
    \label{fig:zcomp}
\end{figure}

Each subject's ITD sensitivity (relative to NH listeners) across all five frequencies ($f_1$ to $f_5$) can be compressed to a single number by calculating the Z-score composite $Z$ (Eq. \eqref{eq:zcomp}), i.e.,
\begin{equation}\label{eq:zcomp}
    Z =\frac{1}{5}\times \sum_{f_i=250~\text{Hz}}^{1250~\text{Hz}}\frac{X_{HI,f_i}-\bar{X}_{NH,f_i}}{\sigma_{NH,f_i}} ,
\end{equation}
where $X_{HI,f_i}$ is the subjects average ITD sensitivity, $\bar{X}_{NH,f_i}$ is the averaged normal hearing sensitivity, and $\sigma_{NH,f_i}$ is the standard deviation of the NH control groups sensitivity at the corresponding frequency $f_i$. The values for NH listeners were taken from \cite{Brughera2013}. The composite Z-scores are shown in Fig. \ref{fig:zcomp}.

The closer the Z-score is to 0, the closer the subjects performance is to the NH values. This representation confirms that P3 performs the best, while P4 performs the worst out of all the subjects. It also shows, that while P4 and P6 have strongly raised thresholds, they are not yet at the maximum value, and are still able to perceive ITDs to some degree. These findings are also in line with the binaural beat experiment from \cite{Calis2021}, where HI subjects showed strong inter-individual differences in their ability to perceive binaural beats. 

\subsection{Discussion}\label{ch:ITDdiscussion}
\begin{figure}[h!]
    \centering
    \includegraphics[width=1\linewidth]{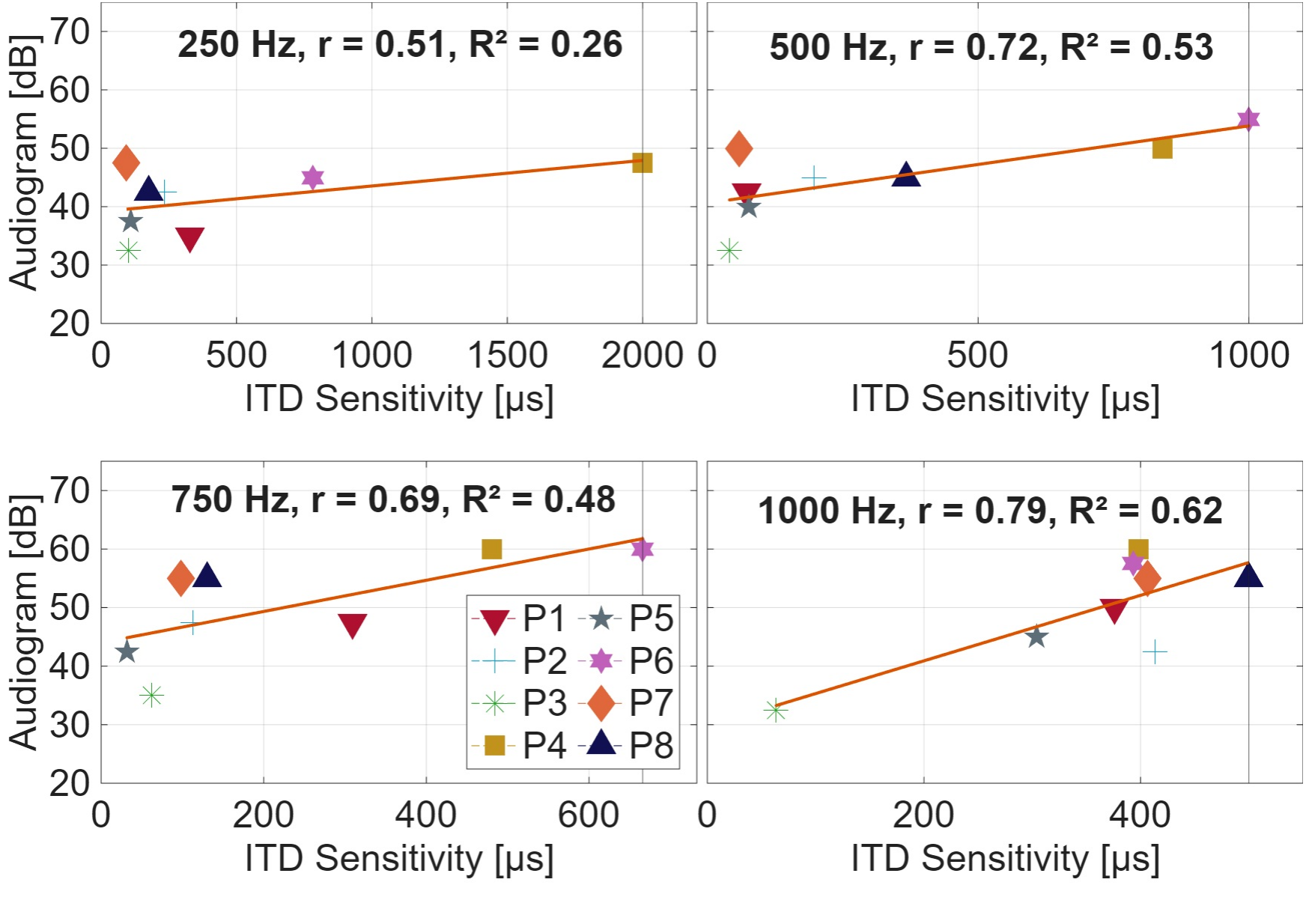}
    \caption{Linear fit (red line) for averaged audiograms between left and right per subject (y-axis) and averaged ITD sensitivity threshold per subject (x-axis) for 250 Hz, 500 Hz, 750 Hz and 1~kHz, as well as the $r$ coefficient of correlation,  $R^2$ coefficient of determination and the maximum possible ITD at that frequency (black line).}
    \label{fig:ITDcorr}
\end{figure}

The ITD thresholds shown in Fig. \ref{fig:ITD_tresholds} show pronounced inter-individual differences. Overall, P3 performs the best with the lowest ITD thresholds. As shown in Fig. \ref{fig:audiograms}, P3 also has the lowest audiometric thresholds out of all the subjects. To further investigate the relationship between the audiometric threshold and the ITD threshold, a linear fit between the data points was calculated, as well as the coefficient of determination $R^2$ using Eq. \eqref{eq:coeff_det}, that is, 
\begin{equation}\label{eq:coeff_det}
    R^2 = 1-\frac{SS_{res}}{SS_{tot}},
\end{equation}
where $SS_{res}$ is the sum of squares of the residuals and $SS_{tot}$ is the total sum of squares. The coefficient of correlation $r$ is then calculated as the square root of $R^2$
\begin{equation}
    r = \sqrt{R^2}.
\end{equation}
The subjects' pure tone audiometric thresholds (averaged between the left and right ear), are plotted over their averaged ITD sensitivity per frequency where both measures are available (250 Hz, 500 Hz, 750 Hz, 1000 Hz) in Fig. \ref{fig:ITDcorr}.

The data shows a weak link at 250~Hz with $r\approx$ 0.51 and $R^2~\approx$~0.257, mainly driven by P4. Towards the higher frequencies, the correlation coefficient increases to values between $r$ = 0.69 and $r$ = 0.79, with 48-62\% of the variance being explained. The approximation of the linear regression also seems to fit well for most subjects, except for P3, who is below the fit line at three out of four frequencies, and P7, who is above the fit line at all four frequencies. In general, the results indicate a strong relation between hearing loss and ITD sensitivity loss. However, it should be noted that 8 subjects are a rather small sample size, which results in large areas without data points. A larger number of subjects might lead to different findings. 

Correlations between hearing level and ITD sensitivity of HI listeners were also investigated in \cite{Spencer2016}. They looked at data at 500 Hz, as well as the averaged values between 250-8000 Hz, and were not able to find a relationship between the subjects hearing loss and their ITD sensitivity, with $R^2$ = 0.023 and $R^2$ = 0.011, respectively. However, they used young HI listeners, where everyone except for one subject was younger than 40 years. 
In another study, \cite{King2014} measured the effects of sensorineural hearing loss and age on TFS and envelope IPD perception. They found significant correlation coefficients between the hearing threshold and the TFS IPDs at 250 Hz and 500 Hz, which is in line with our findings. 

\section{Experiment 2: Speech Reception Thresholds}\label{ch:SRTs}
After assessing the subjects' impairment in ITD perception, the second experiment focused on the effect of the ITD-to-ILD transformation method and the separate binaural cues on speech intelligibility. As a measurement metric, the 50\% speech reception threshold ($\text{SRT}_{50}$) was chosen. The German matrix test OLSA \citep{Wagener1999} was used for the test procedure and the target speaker speech material.

Subjects were presented two interfering speakers from an audiobook at either $\pm 60^\circ$~with a \emph{central} target speaker at $0^\circ$, or with interfering speakers at $-60^\circ~\text{and}~0^\circ$ for a \emph{lateral} target speaker at $60^\circ$, like in \cite{Bäumer2025}. It was ensured, that the audiobook sentences were longer than the target OLSA sentences, and that they did not contain long gaps between words. To spatialize the different speakers, the KEMAR HRTFs from the Aachen database \citep{Braren2020} was used.

\subsection{Half-Gain Rule Filter}
\begin{figure}[h!]
    \centering
    \includegraphics[width=1\linewidth]{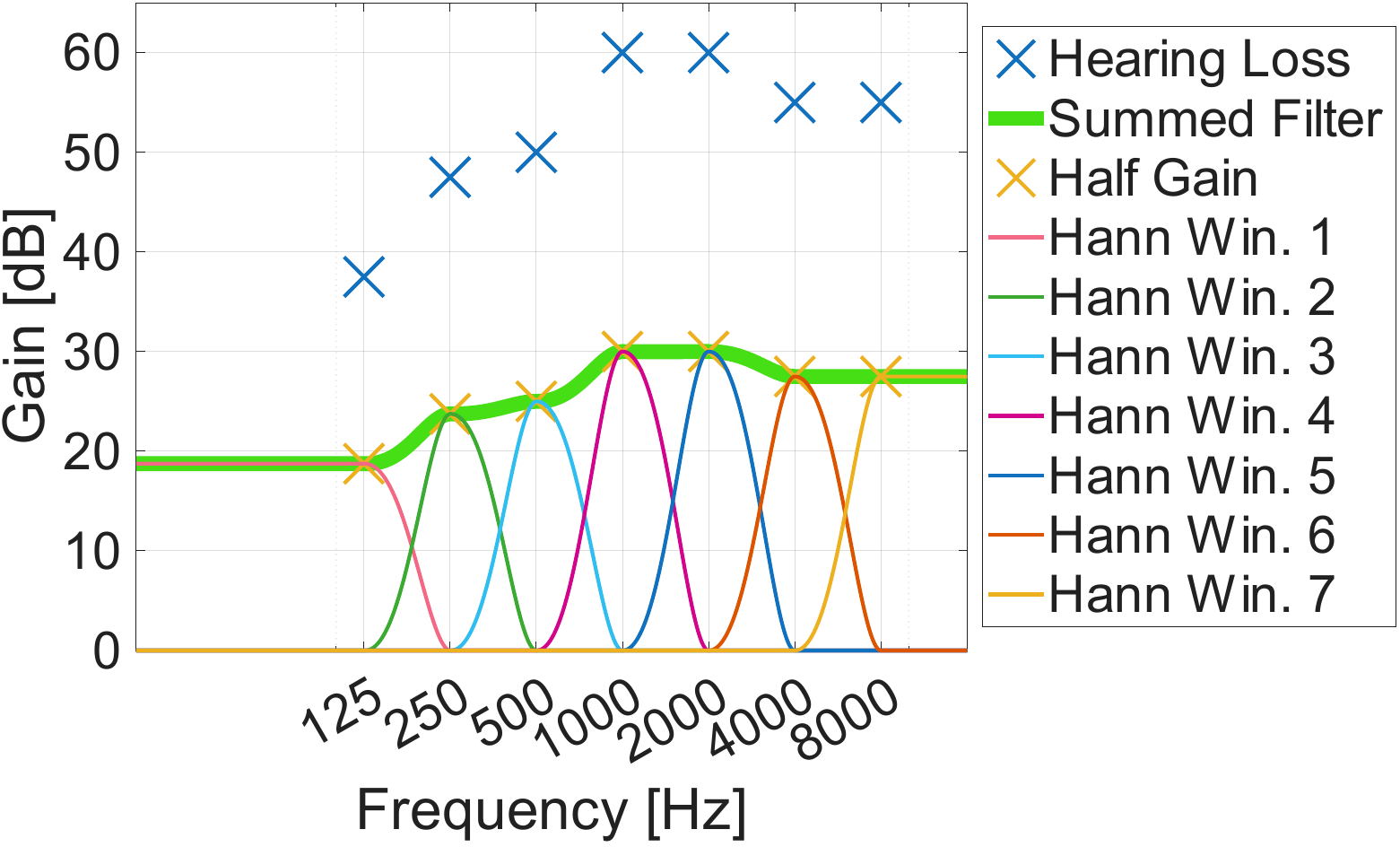}
    \caption{Half-gain rule filter built from overlapping Hann-windows. The blue crosses denote the subjects hearing loss based on the audiogram (averaged between left and right), the orange crosses denote the corresponding gain the filter should have at that frequency, and the different Hann windows overlap to produce the desired magnitude filter response (thick green line).}
    \label{fig:half_gain}
\end{figure}

The presented signals had a general starting level of 65~dB~SPL. To compensate for the participants individual hearing loss, a half-gain rule \citep{Lybarger1963} filter built from overlapping, octave frequency band wide Hann windows was applied to the stimuli to boost the level in the respective frequency bands. An example of such a filter is shown in Fig. \ref{fig:half_gain}. 

The gains for each Hann window were taken from the participants audiogram by halving the hearing loss (Fig. \ref{fig:audiograms}). The resulting frequency response was then used to design an arbitrary magnitude filter of order $N~=~300$ with a linear phase response, resulting in individually different stimuli and stimulus levels for each participant. In order to not risk additional hearing damage and to stay away from the subjects uncomfortable level thresholds (UCL), the stimuli were limited to a maximum level of 85~dB~SPL after the filter and during the SRT measurement.

\subsection{Procedure}
To familiarize subjects with the measurement procedures, in the beginning they listened to the \emph{unprocessed} condition with a central target speaker. At first they were presented with the target speaker only, without interfering speakers so the subjects knew which voice is the target speaker. Afterwards, the interferers were added with decreasing SNR to present different conditions the subjects might encounter, ranging from +12 to -10 dB SNR. Lastly, the subjects were presented with the \emph{unprocessed colocated} condition, where both the target and the interferers were positioned in the center in front of the listener. In case subjects were unable to distinguish between target and interferers (even though the voices were different), there was an option to switch to female interferers instead, but all subjects reported their ability to differentiate between the voices. 

The actual measurements were presented with a starting SNR of 5 dB, which was above threshold for all the participants. In total, the participants listened to 11 different measurement conditions (described in the next section). The 11 different measurement conditions were presented in random order, with mandatory breaks after every 3 or 4 conditions, to ensure the subjects remain concentrated. 

\subsection{Binaural conditions}\label{sec:configurations}

In the listening experiment, the KEMAR HRTFs were manipulated to yield five different HRTF conditions. 

As a baseline, the first condition used of the \emph{unprocessed} HRTF, which was not manipulated and, thus, contained the natural low-frequency ITDs and high-frequency ILDs for the KEMAR.

In the second condition, the ITDs from the \emph{unprocessed} HRTF were removed by unwrapping the phase of the left and right channel for each horizontal azimuth DOA and frequency bin, and replacing it with the averaged value between left and right. This removed the Interaural Phase Differences (IPDs) and, thus, the ITDs, while leaving the magnitude spectrum unchanged. This condition is referred to as \emph{noITD} and is supposed to simulate severe spatial hearing loss by entirely removing the ability to process ITDs. The removed ITDs can also be seen in \ref{fig:ITDs} b), as the black line at 0~µs.

To investigate the influence of the ILDs, in the third condition the ILDs were removed by equalizing the HRTF's left and right power spectra using Eq. \eqref{eq:removeILD}+\eqref{eq:removeILD2} per frequency bin $k$ for each horizontal azimuth DOA. The \emph{unprocessed} phase spectrum is unaltered so that the ITDs are still present. This condition is referred to as \emph{noILD}.

\begin{flalign} \label{eq:removeILD}
    L'(k) = \frac{L(k)}{|L(k)|}\sqrt{\frac{1}{2}(|L(k)|^2+|R(k)|^2)}, \\
    R'(k) = \frac{R(k)}{|R(k)|}\sqrt{\frac{1}{2}(|L(k)|^2+|R(k)|^2)}
    \label{eq:removeILD2}
\end{flalign}

In the fourth condition, the proposed transformation method was applied to the \emph{noILD}-HRTF, substituting the absent ITDs with the low-frequency ILDs, to get the \emph{$\text{transformITD}_{\text{sub}}$} condition. These four binaural conditions were also measured in \cite{Bäumer2025}.

To further test if residual ITD processing in hearing impaired listeners can be used to get an even higher binaural advantage, in the fifth condition the ITD transformation method was applied to the \emph{unprocessed} HRTF, adding the low-frequency ILDs while the ITDs are still present. This is also closer to how a beamformer in a hearing aid would operate, since in the implementation of the transformation method, ITDs should not be removed beforehand to allow the listener to make use of residual ITD processing. This final condition is referred to as \emph{$\text{transformITD}_{\text{add}}$}.

All of the HRTFs were rendered using the Room Acoustic Simulation Toolbox RAZR \citep{Wendt2014} in Matlab \citep{Matlab} to obtain the spatialized Binaural Room Impulse Responses (BRIRs) for the source and receiver positions. In RAZR, an anechoic room was designed by setting the order of Image Sources to 1, disabling the Feedback Delay Network and setting the absorption coefficients of each wall to 0.99. These BRIRs were then used to create the 11 measurement conditions, five with a central speaker, five with a lateral speaker, and a colocated one. 

\subsection{Results}\label{ch:SRTresults}
\begin{figure}[h!]
    \centering
    \includegraphics[width=1\linewidth]{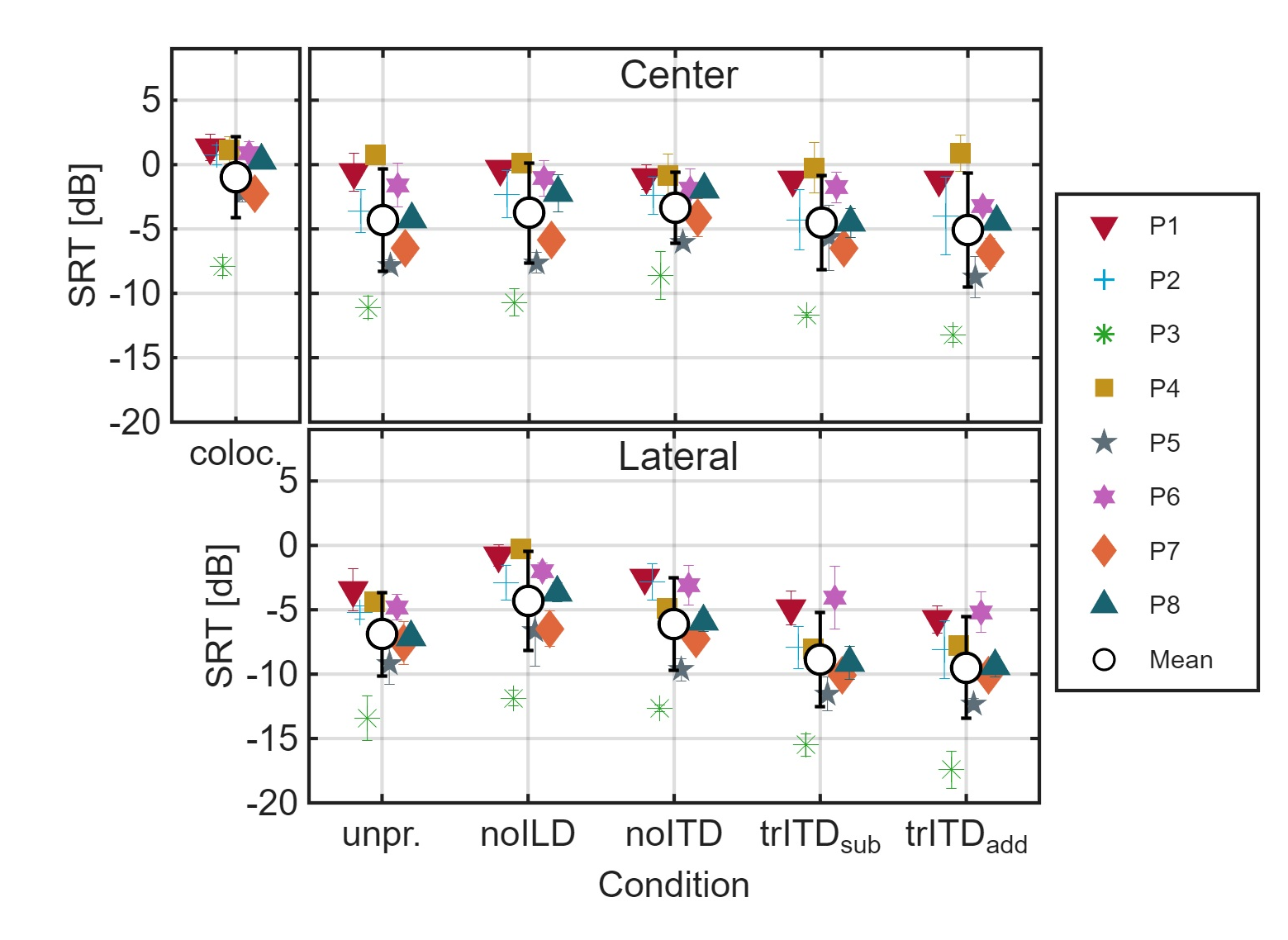}
    \caption{Speech Reception Thresholds of the Hearing Impaired subjects P1 to P8 in the different HRTF and spatial conditions. The coloured filled symbols denote the individual subjects performances, the errorbars indicate the standard deviation after averaging the three measurements and the empty circle denotes the averaged value across all subjects, with the standard deviation.}
    \label{fig:srt_overall}
\end{figure}

The SRTs for each subject, averaged across all three measurement repetitions, are shown in Fig. \ref{fig:srt_overall}, along with the mean value across all subjects. 

\begin{figure}
    \centering
    \includegraphics[width=1\linewidth]{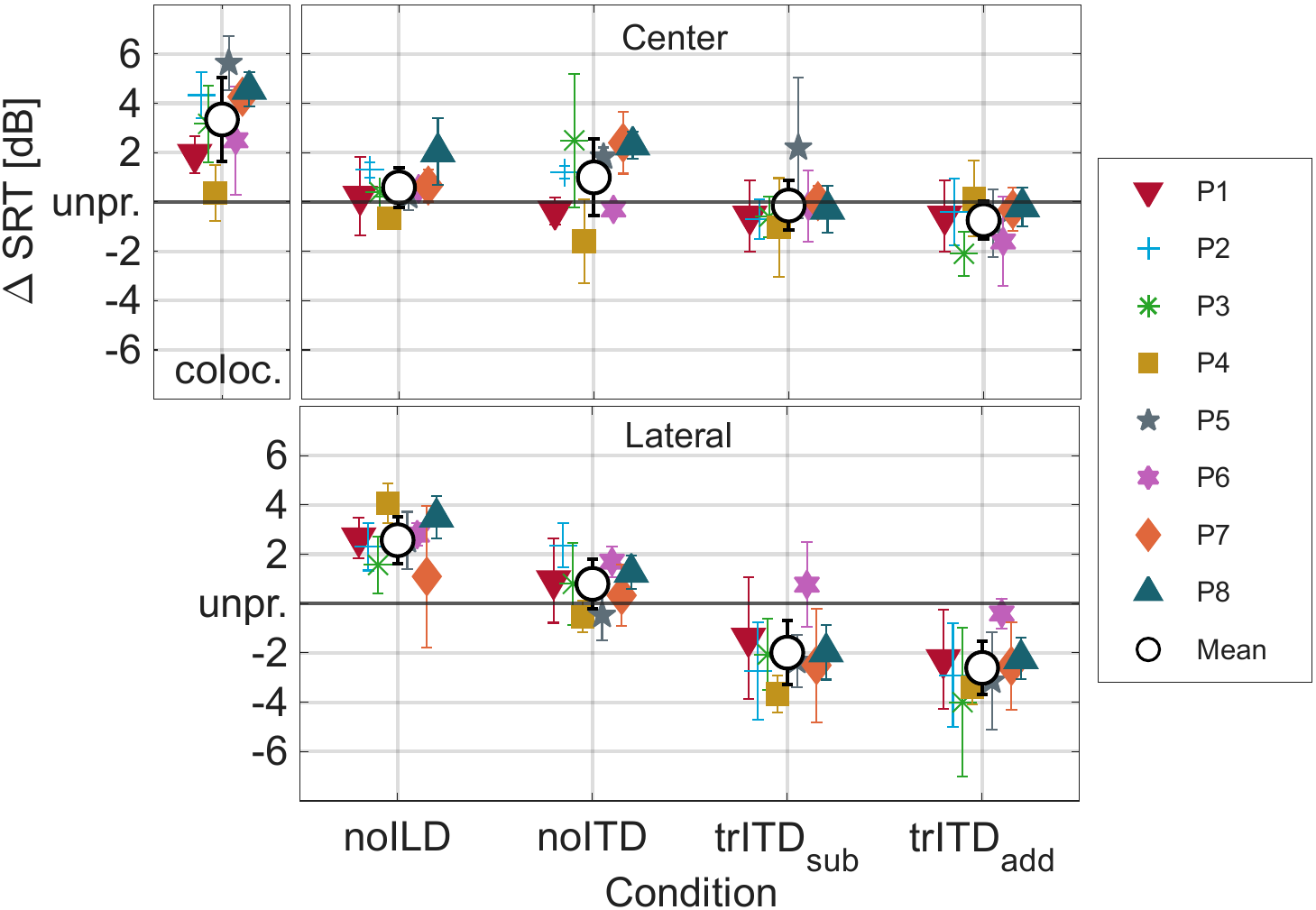}
    \caption{$\Delta$ Speech Reception Threshold reduction of the Hearing Impaired listeners relative to the unprocessed condition as a baseline for the different HRTF and spatial conditions. The coloured filled symbols denote the individual subjects performances, the errorbars indicate the standard deviation after averaging the three measurements and the empty circle denotes the averaged value across all subjects, with the standard deviation.}
    \label{fig:srt_relative}
\end{figure}

In general, it can be seen that the subjects perform better in the lateral conditions compared to the central conditions, as shown by the lower SRTs. However, within each condition, the subject's performances deviate strongly from one another. For a central target speaker in the \emph{$\text{transformITD}_{\text{add}}$} condition, the difference between the best and worst performing subject is about 14~dB~SRT. In order to better compare the effect of the different HRTF conditions, Fig. \ref{fig:srt_relative} shows the subjects SRTs relative to their \emph{unprocessed} condition SRTs as a baseline for the central and lateral target speaker respectively. 

In this representation, the differences between the subjects within a condition reduce to a maximum difference of about 6 dB. For the \emph{colocated} (relative to the central \emph{unprocessed} condition), it is visible that the subjects (on average) achieved SRM of about 3.5 dB. Removing the ILDs in the central case only decreased the SRTs (and thus, the speech intelligibility) by about 0.5 dB. Removal of the ITDs was, on average, more detrimental, with a reduction of about 1 dB. However, this includes a much larger standard deviation between the subjects, than in the ILD case. The individual subjects' results also show, that more subjects are negatively affected by ITD removal than by ILD removal in the central case. The biggest decrease in SRTs after removing the ITDs was found in P3, P5 and P7, which are also the subjects with the best ITD thresholds, as shown in Fig. \ref{fig:zcomp}. 

When substituting the missing ITDs with low-frequency ILDs (\emph{$\text{trITD}_{\text{sub}}$}), on average, a small improvement of ca. 0.1~dB over the unprocessed condition is observed in the central case. By preserving the ITDs in addition to the ITD-to-ILD-transformation (\emph{$\text{trITD}_{\text{add}}$}), the SRTs improve further by about 0.7 dB. 

In the lateral case (Fig. \ref{fig:srt_relative}, right side), the importance of ITDs and ILDs seems to switch, with the \emph{noILD} condition decreasing the SRTs by ca. 2.6 dB and the \emph{noITD} condition decreasing the SRTs by only 0.8 dB. Both of the transformations, however, lead to a much larger benefit in the lateral conditions than for the center one. The transformation without residual ITDs (\emph{$\text{trITD}_{\text{sub}}$}) improves the SRTs (i.e. the speech intelligibility) on average by about 2 dB, and by preserving the ITDs (\emph{$\text{trITD}_{\text{add}}$}) the performance benefit increases to about 2.6 dB. The highest improvement overall was found in P3 in the lateral \emph{$\text{trITD}_{\text{add}}$} condition, with an improvement of 4 dB.

\begin{figure}[h!]
    \centering
    \includegraphics[width=1\linewidth]{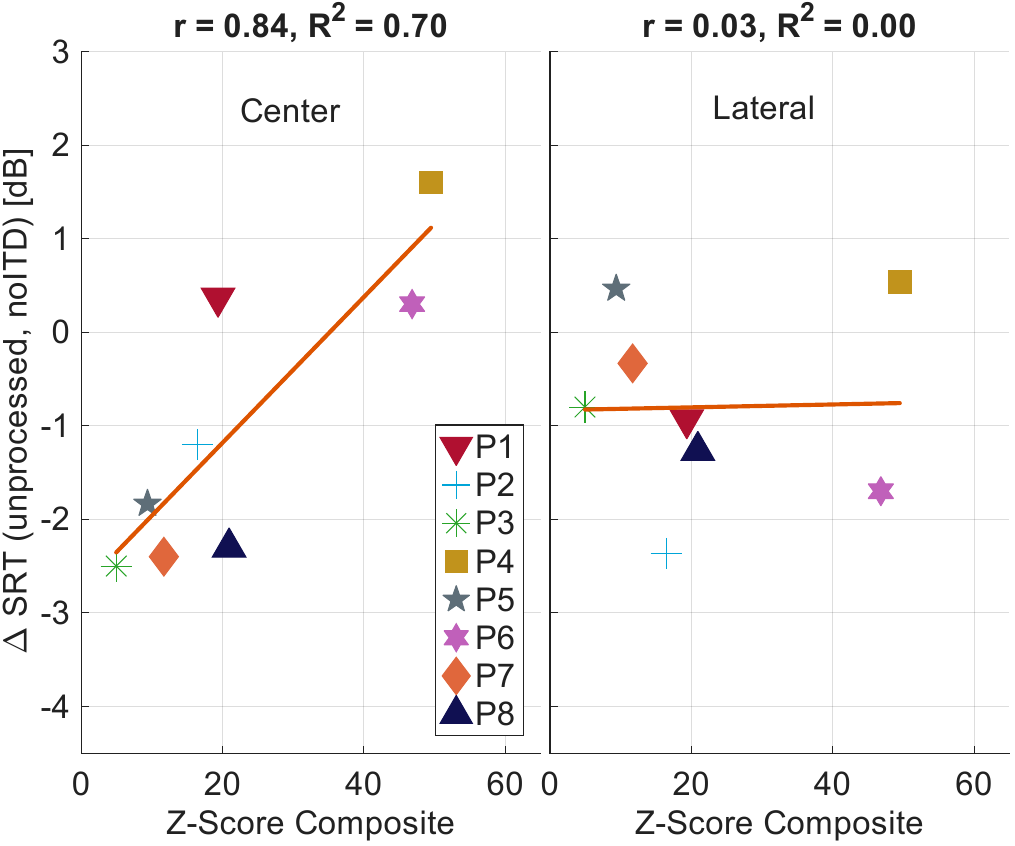}
    \caption{Correlation between the reduction in SRTs due to removal of ITDs (difference between \emph{unprocessed} and the \emph{noITD} condition for the central and lateral target, and the composite Z-scores, representing the ITD sensitivity).}
    \label{fig:corr_noITD}
\end{figure}

To further emphasize the connection between ITD sensitivity (in the form of composite Z-scores, Eq. \eqref{eq:zcomp}) and the reduction in SRTs between the \emph{unprocessed} and the \emph{noITD} condition, the relationship between them is shown in Fig. \ref{fig:corr_noITD}.

In the central case, a high correlation of $r=0.83$ is visible, indicating a close link. In the lateral case, however, ITDs seem to be less important. This is in line with the findings in Fig. \ref{fig:srt_relative} and in \cite{Bäumer2025}, where the \emph{noILD} condition in the lateral case is much more detrimental than the \emph{noITD} condition, indicating a stronger relative importance of the ILDs due to the better-ear effect. 

To test the statistical significance of the findings, a two-way repeated measures ANOVA was conducted on the factors "Condition" and "Repetition". The factor "Condition" was found to be very significant ($F$~=~48.89, $p$~$<$~0.001), the factor "Repetition" was found to be significant ($F$~=~3.92, $p$~=~0.043) and the interaction between them was not ($F$~=~1.62, $p$~=~0.056). A further post-hoc analysis was conducted on the individual conditions by calculating pairwise t-tests for the conditions relative to the unprocessed baseline (as shown in Fig. \ref{fig:srt_relative}). After applying the Benjamini-Hochberg correction \citep{Benjamini1995}, it was found that in the central case, only the \emph{colocated} ($p$~=~0.0016) and the \emph{$\text{transformITD}_{\text{add}}$} condition ($p$~=~0.0411) are significant. The \emph{noILD} ($p$~=~0.0886), \emph{noITD} ($p$~=~0.1297) and \emph{$\text{transformITD}_{\text{sub}}$} condition ($p$~=~0.7143) are not. In case of the lateral target speaker, the \emph{noILD} ($p$~=~0.0004), \emph{$\text{transformITD}_{\text{sub}}$} ($p$~=~0.0056) and \emph{$\text{transformITD}_{\text{add}}$} condition ($p$~=~0.0006) were found to be significant. Only the \emph{noITD} condition was not ($p$~=~0.0779).

\section{\label{sec:discussion} Discussion}
The present results show a significant improvement in the SRTs when applying the ITD-to-ILD transformation while preserving the ITD cues (\emph{$\text{transformITD}_{\text{add}}$}) compared to the \emph{unprocessed} condition. Subjects improved on average by about 2.6~dB for a lateral target speaker, while the largest improvement was found in P3 with 4~dB. Interestingly, for the same condition, we also found a slight improvement in SRTs for a central target speaker. Since the ITDs in the center are zero, the transformation only affects the interfering speakers, but still causing an improvement in speech intelligibility. This could indicate that the improvement in speech intelligibility is due to enhanced lateralization of the interfering speakers, resulting in higher SRM due to improved source segregation. Another explanation might lie in the implementation of the transformation method, where the ipsilateral ear is boosted and the contralateral ear is attenuated. If the interferer on one side is quiet in a glimpse, then the other interferer will be boosted in the ipsilateral ear, but attenuated in the contralateral ear. This could enhance the better-ear listening effect, with the contralateral ear having the better SNR in this glimpse. Thus, interestingly, since in this case no direct processing is done on the target speaker in the center, we can infer that the ITD-to-ILD transformation is able to improve speech intelligibility in all directions, no matter where the target speaker is.

Comparing the SRM between the NH and the HI listeners, we can see that the HI listeners only benefit by about half as much (3.5~dB here vs ca. 7~dB in \cite{Bäumer2025}). This reduction is to be expected due to the HI listeners reduced access to the binaural cues, and also shown in literature (e.g. \cite{Marrone2008}). In case of the \emph{noITD} condition for the NH and HI subjects, we see that in the central case, ITD removal is more detrimental to NH listeners (ca. 2~dB), which could be explained by the subjects reduced ITD sensitivity. In the lateral case, however, ITD removal is more detrimental to HI listeners (ca. 1~dB). The fact that there is an SRT reduction in the HI listeners shows, however, that simply removing the ITDs from the signals entirely (as was done in \cite{Bäumer2025}) is not an accurate simulation of spatial hearing loss, since there is still some degree of ITD processing in the HI listeners. This is also supported by the fact, that the subjects perform better in the \emph{$\text{transformITD}_{\text{add}}$} condition, where ITDs are preserved, than in the \emph{$\text{transformITD}_{\text{sub}}$} condition, where ITDs are removed. Similarly, \cite{Best2019}, also suggested that HI listeners with poor pure-tone ITD sensitivity still benefit from ITDs found in the envelopes of speech signals. 

Looking at the individual results of some of the subjects, there are some interesting outliers and phenomenons. Participant P4, which showed the highest ITD thresholds in the first experiment (e.g. Fig. \ref{fig:zcomp}) showed an improvement in speech intelligibility of almost 2~dB when the ITDs were removed from the signals (for the central target speaker). The reason for this finding is not clear. One possible explanation would be that the ITD information that is still present in the stimuli somehow gets distorted in the subject's auditory system, causing a decrease in speech intelligibility. This theory is further supported by the fact, that P4 performs better in the central \emph{$\text{transformITD}_{\text{sub}}$} condition than in the \emph{$\text{transformITD}_{\text{add}}$} condition, though the performance in both of these conditions is worse than in the \emph{noITD} condition.
For P5, the averaged data across all three measurements shows an SRT decrease of about 2~dB in the central \emph{$\text{transformITD}_{\text{sub}}$} condition, compared to the \emph{unprocessed} condition. The individual measurements, however, show an effect of training where in the first measurement the performance compared to the \emph{unprocessed} condition was impaired by about 5~dB, in the second measurement was impaired by 2.3~dB and in the third measurement there was an improvement of 1.1~dB. This indicates an effect of training for this rather unnatural type of acoustic manipulation, where subjects first need to get used to the different spectrum. This is also backed by the results of the two-way repeated measures ANOVA, that show a significant effect of the repetition.

In general, the results we found in this study (regarding the improvement in speech intelligibility) are well in line with previous ITD-to-ILD transformation methods, such as \cite{Brown2014, Richardson2025, Bäumer2025}. It remains to be seen, however, how this transformation alters the subjects ability to localize sounds, since manipulations of the binaural cues could lead to an unnatural sense of sound localization. It could also be interesting to include more low-frequency bands when using another HRTF set or an actual cochlear implant algorithm, to not be limited to extrapolating below 400 Hz. More frequency bands (in the 250 to 400 Hz range) with actual ITD estimation for the transformation might increase the benefit of the ITD-to-ILD transformation.

To use the transformation method we propose here in an applied setting requires some more work. As of now, the method relies on oracle knowledge in the form of measured HRTFs and perfectly defined directions of arrival. A more application-oriented approach using a beamformer with acoustic transfer function (ATF) estimation that applies the same ITD-to-ILD transformation as proposed here is presented in \cite{DeVries2025}.

\section{BSIM Predictions}
To test the plausibility of the SRT measurements, the auditory model BSIM \citep{Beutelmann2010} was used to predict SRTs in the same target and interferer scenarios. To ensure the models ability to predict the SRTs when using manipulated HRTFs, it was first tested on the NH data presented in \cite{Bäumer2025}. When calibrated on the colocated SRT (averaged across all participants) the model performed well in all different conditions. The largest deviation of the model was in the \emph{$\text{center transformITD}_{\text{sub}}$} condition with an overprediction of 1.9 dB, which is still in the measured standard deviation. The smallest deviation is in the \emph{lateral unprocessed} condition, with a difference of 0.2~dB. Across all conditions, the model has a mean deviation of only 0.78 dB SRT, meaning it is well able to predict the different binaural manipulations. To extend this investigation to HI listeners, the BSIM can be personalized by including individual audiograms as internal masking noise.

\begin{figure}[h!]
    \centering
    \includegraphics[width=1\linewidth]{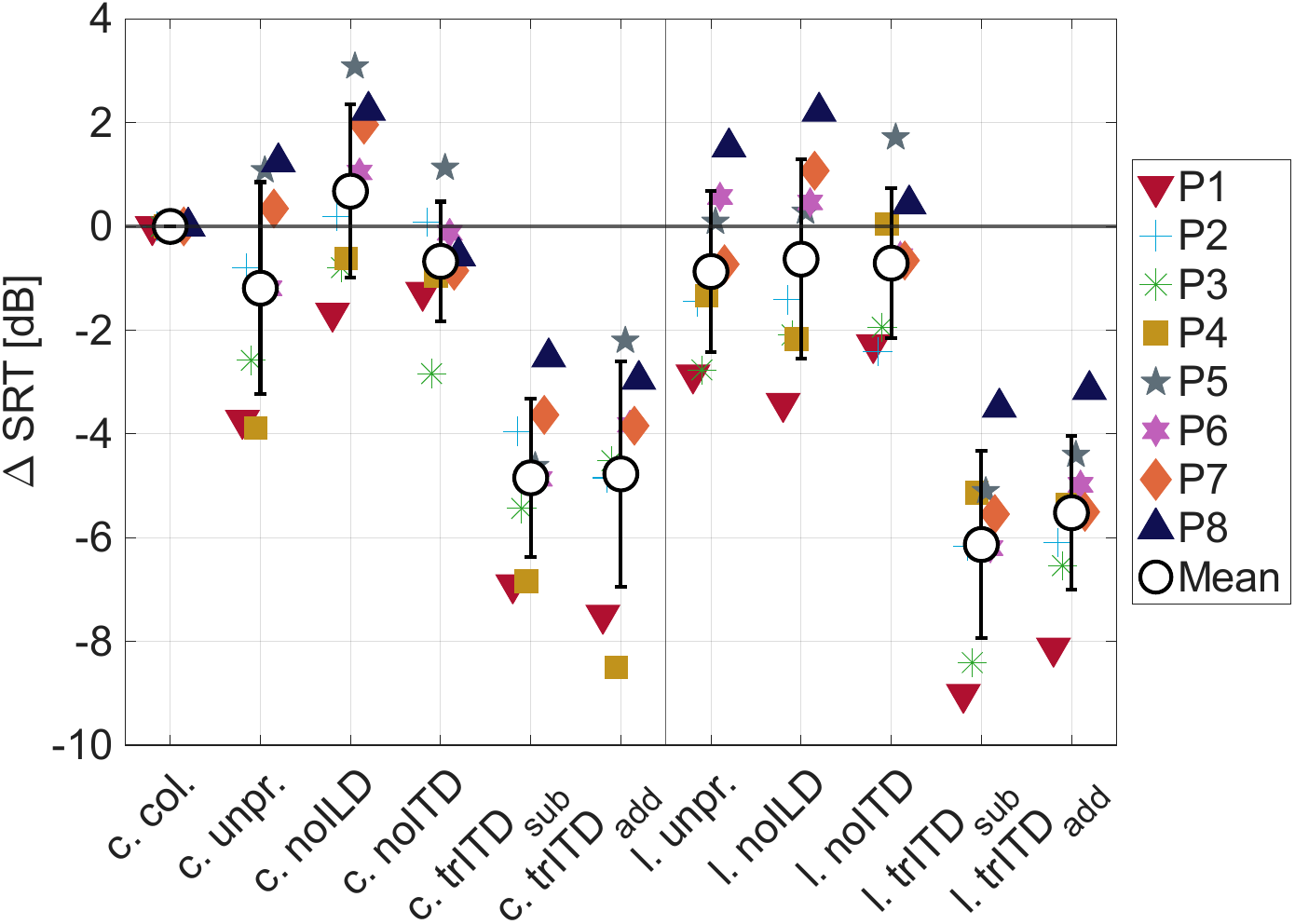}
    \caption{Difference between BSIM prediction and measured SRTs for each subject and measurement condition.}
    \label{fig:bsim_diff}
\end{figure}

To obtain personalized BSIM predictions, the model was applied independently for each participant in the listening test. After equalizing all input sources to 65~dB~SPL, the half-gain rule, as described in Fig.~\ref{fig:half_gain}, was applied following each subject's audiogram similar to the preparation of the stimuli in the listening experiment. The audiogram was also provided as input to BSIM, which is converted to additional masker noise of the appropriate level. The reference SII values were calibrated to match the mean across the three repeated measurements of the individual's measured \emph{colocated} SRT, and are given in Table~\ref{tab:bsim_sii}. The resulting SRT prediction errors (measured SRTs - predicted SRTs) are shown in Fig.~\ref{fig:bsim_diff}.

\begin{table}[h!]
\centering
\begin{tabular}{|l|l|l|l|l|l|l|l|}
\hline
   \textbf{P1} & \textbf{P2} & \textbf{P3} & \textbf{P4} & \textbf{P5} & \textbf{P6} & \textbf{P7} & \textbf{P8} \\ \hline
 0.11        & 0.12        & 0.03        & 0.10        & 0.07        & 0.08        & 0.03        & 0.06        \\ \hline
\end{tabular}
\caption{SII$_{\text{ref}}$ values used in the BSIM model for each participant}
\label{tab:bsim_sii}
\end{table}

First of all, there is more uncertainty in validating the model outputs compared to using NH data, as the individualization of the model leads to less data points being available per prediction. This is evident in the large inter-individual variability of prediction errors in Fig.~\ref{fig:bsim_diff}, spanning both under- and overpredictions in most conditions. Overall, the model predictions seem a little less accurate than for the averaged NH case using data from \cite{Bäumer2025}. The closest matches are found for P6, and the most accurately predicted condition is the central \emph{noITD} condition. For some subjects, systematic under- (e.g. P1) or overpredictions (e.g. P8) can be observed. These systematic errors can partially be explained by the model being calibrated to a measured \emph{colocated} SRT with high variance/uncertainty. Any erroneous offset in this calibration would lead to an offset in all subsequent predictions.

The \emph{unprocessed} conditions show significant prediction errors for some subjects --- especially compared to the respective NH predictions --- despite the BSIM being designed for this task and providing a method for incorporating personalized HI. The \emph{noILD} and \emph{noITD} conditions have a slightly lower mean prediction error, though the inter-individual variability is similar. In both \emph{transformITD} conditions, on the other hand, the SRTs are extremely underpredicted for every subject. Further tests showed that this effect is much less pronounced when leaving out the individualized audiogram inputs. As the other conditions' predictions do improve when including audiograms, this suggests that the specific internal masking noise mechanism that models hearing thresholds is only effective for `natural' binaural cues, not artificially enhanced binaural cues.

\begin{figure}[h!]
    \centering
    \includegraphics[width=1\linewidth]{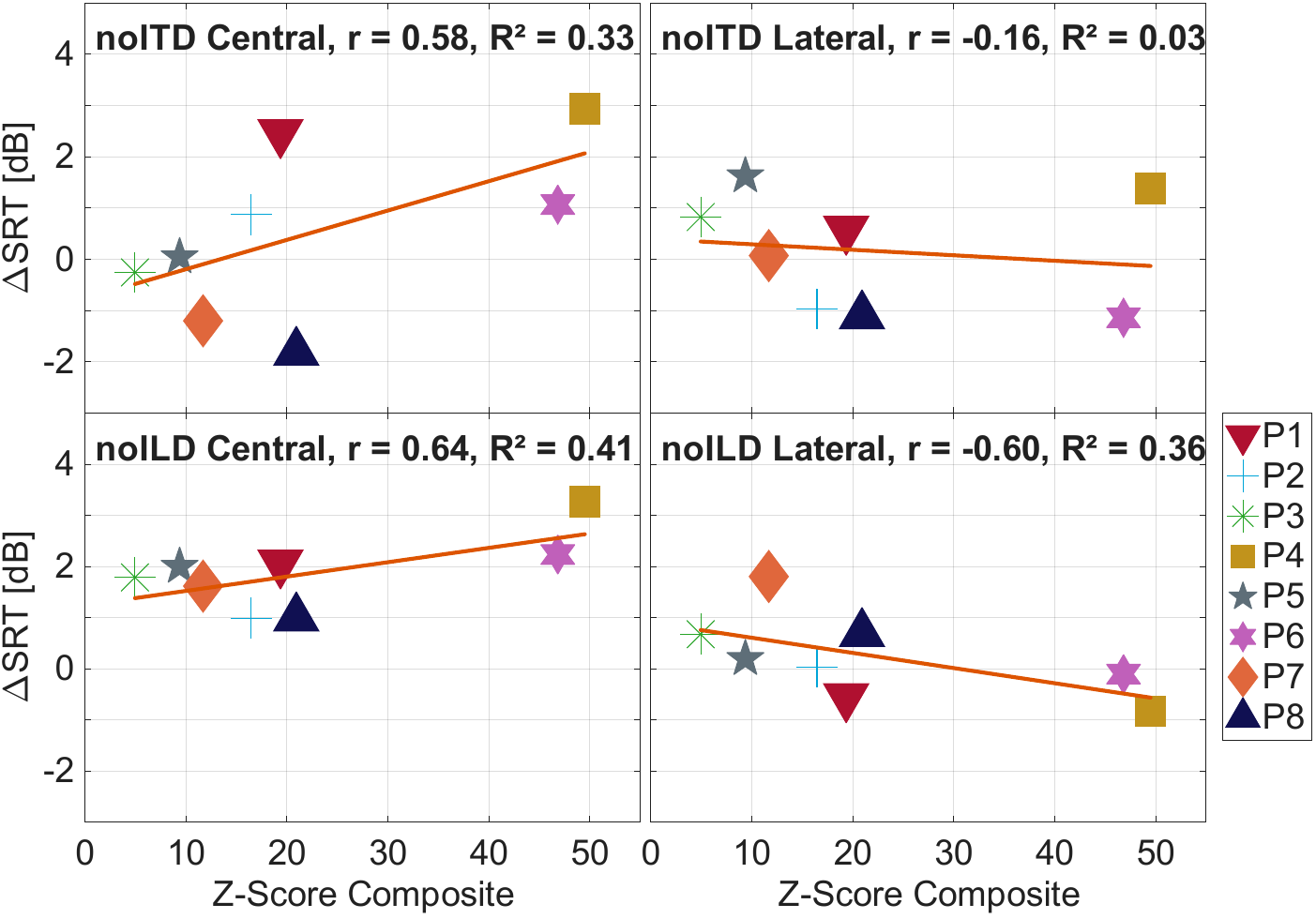}
    \caption{Change in prediction error between BSIM predictions for the \emph{noITD} and \emph{noILD} condition compared to the prediction error in the \emph{unprocessed} condition, correlated to the ITD sensitivity in the form of the Z-score composite per subject. }
    \label{fig:corr_rmCues}
\end{figure}

To investigate the importance of ITD sensitivity to the SRT contributions of individual binaural cues, Fig.~\ref{fig:corr_rmCues} shows linear regressions between composite Z-scores (Eq. \eqref{eq:zcomp}) and the difference between measured and predicted SRT gaps between \emph{unprocessed} and either \emph{noILD} or \emph{noITD} conditions (i.e., whether the effect from removing a cue is under- or overpredicted). In the central conditions, moderate positive correlations exist, suggesting that subjects with ITD sensitivity worse than NH listeners (i.e., high Z-scores) tend to have the effects of removing cues exaggerated in the BSIM. In the lateral cases, there is only a moderate but negative correlation for the \emph{noILD} condition. For a majority of subjects, predicting \emph{unprocessed} SRTs with the \emph{noITD} model would improve performance, although this seems only weakly correlated to composite Z-scores. It should be investigated whether including an ITD sensitivity measure (like those of Fig.~\ref{fig:ITD_tresholds}) as input to the BSIM might improve personalized HI predictions. No significant model output correlations have been found with respect to the audiograms or to left--right audiogram imbalance (at least for the mild imbalances included in this research).

\section{\label{sec:conclusion} Conclusion}
In this paper an ITD-to-ILD transformation method was proposed, with the aim of restoring the binaural benefit in speech intelligibility for hearing impaired listeners, which could be partially lost due to insensitivity to ITDs. Previous work using the same method \citep{Bäumer2025} has shown that substituting removed ITDs with low-frequency ILDs can nearly completely restore the lost binaural benefit in normal hearing listeners for a central target speaker, and even improve performance by about 2~dB compared to the baseline for a lateral target speaker. In this work, the same experiment was conducted with hearing impaired listeners, with an additional experiment to measure their sensitivity to ITDs. It was found that all hearing impaired subjects had different degrees of ITD perception deficits, with two subjects having strongly increased thresholds in each test frequency (250 Hz to 1250~Hz), some subjects having barely increased thresholds in the lower test frequencies but all subjects being less sensitive at 1250~Hz. The ITD sensitivity thresholds also showed a good amount of correlation with the subjects audiograms per frequency up to $r = 0.79$. 

Applying the ITD-to-ILD transformation while preserving the ITDs in the signal increased the speech intelligibility for target speakers in all directions, by up to 4 dB (with two interfering speakers and the target at $60^\circ$). Despite the subjects reduced sensitivity to ITDs, preserving the ITDs still yielded higher SRT improvements than using the transformation method without preserving the ITDs. Speech intelligibility also improved for a central target speaker, where the transformation method only influenced symmetrically placed interfering speakers. This shows, that removing the ITDs completely, as was done in \cite{Bäumer2025}, is not an accurate way to simulate spatial hearing loss. In general, these findings suggest that this method, when implemented in hearing aids or binaurally implanted cochlear implants, could be beneficial to increase speech intelligibility in hearing impaired listeners.

\begin{acks}
 The authors thank the Hörzentrum Oldenburg gGmbH for providing the participants for the experiments from their database, and thank the participants for their time and effort to take part in the study.
\end{acks}

\section*{Statements and Declarations}
\subsection*{Ethical considerations}
The Commission for Research Impact Assessment and Ethics of the Carl von Ossietzky Universität Oldenburg had no objection against the measurement procedures (approval Drs.EK/2021/073-02).
\subsection*{Consent to participate}
The participants of the study confirmed their consent to participate in a written form.
\subsection*{Consent for publication}
The participants of the study confirmed their consent for publication in a written form.
\subsection*{Declaration of conflicting interest}
The authors declared no potential conflicts of interest with respect to the research, authorship, and/or publication of this article
\subsection*{Funding statement}
This research was funded by the Deutsche Forschungsgemeinschaft (DFG, German Research Foundation) under Germany's Excellence Strategy – EXC 2177/1 - Project ID 390895286 and partly supported by the Dutch Research Council (NWO).
\subsection*{Data availability}
Due to lack of ownership of the HRTF and the OLSA material, it is not possible to share any codes. The participants listening test results (ITD thresholds and SRTs) are available

\bibliographystyle{SageH}
\bibliography{my_bib.bib}

\end{document}